\documentclass{elsart}

\usepackage{graphicx,amssymb}

\journal{Physics Letters A}

\usepackage{psfrag}

\begin{document}

\begin{frontmatter}

\title{Inferring nonlinear parabolic field equations from modulus
data}

\author{Rotha P. Yu\corauthref{cor1}},
\ead{Rotha.Yu@spme.monash.edu.au}
\corauth[cor1]{Corresponding author}
\author{David M. Paganin},
\author{Michael J. Morgan}

\address{School of Physics and Materials Engineering, Monash
University, Victoria 3800, Australia}

\begin{abstract}
We give a means for measuring the equation of evolution of a
complex scalar field that is known to obey an otherwise
unspecified (2+1)-dimensional dissipative nonlinear parabolic
differential equation, given field moduli over three
closely-spaced planes. The formalism is tested by recovering
nonlinear interactions and the associated equation of motion from
simulated data for a range of (2+1)-dimensional nonlinear systems,
including those which exhibit spontaneous symmetry breaking. The
technique is of broad applicability, being able to infer a wide
class of partial differential equations, which govern systems
ranging from nonlinear optics to quantum fluids.
\end{abstract}

\end{frontmatter}

The laws of physics profoundly simplify an otherwise bewildering array of data collected by our measuring instruments \cite{Wheeler1982}. Many of these laws take the form of partial differential equations governing the spatial and temporal evolution of field quantities associated with radiation or matter. These equations involve quantities that are often either single-component or multi-component complex fields. One may enquire whether it is
possible to uniquely determine the associated equation of evolution from measurements made on such fields.  In this context, there has been much research on such ``identification'' \cite{Nelles2001} of partial differential equations, both linear and nonlinear, given measurements of the field itself (see e.g., \cite{Voss1998,Bar1999,Gal2003}). However, this earlier work restricts consideration to intrinsically real field quantities which are sufficiently slowly varying that both their amplitude and phase may be directly measured, providing data that can subsequently be used to determine the equation governing the evolution of the
field.
\par
Such existing methods are not applicable to complex fields whose phase is not known, whether these be intrinsically complex fields or rapidly-oscillating real fields described by a complex analytic signal \cite{Gabor1946}. In the case of complex quantum-mechanical fields, the phase is not directly measurable -- however, following a line of investigation which dates at least as far back as Pauli \cite{Pauli1933}, methods do exist for inferring wavefunction phase from non-interferometric measurements of probability density \cite{Aharonov1993,AllenOxley2001,Paganin2001,Tan2003}. Regarding the case of classical fields such as scalar electromagnetic disturbances, limitations of present detector technology imply that the phase of the disturbance is not directly measurable, at optical and higher frequencies \cite{Born1993}; following Gabor, such real fields are conveniently described using their complex analytic signal, which is so called on account of the fact that
extension to complex time yields a function of a complex variable that is analytic in the lower half of the complex plane \cite{Gabor1946}.
\par
In the present paper, we demonstrate the feasibility of
determining the partial differential equations that govern the
evolution of complex fields, given modulus data alone.  This
comprises a means for ``measuring'' certain partial differential
equations, which govern complex fields whose modulus is known but
whose phase is unknown.
\par
Consider a complex (2+1)-dimensional scalar field $\Psi$, governed
by an equation belonging to the following class of dissipative,
nonlinear parabolic equations \cite{Tan2003}:
\begin{eqnarray}
    \left[i\alpha\frac{\partial}{\partial z}+\nabla^2_{\perp}
    +f(|\Psi|)+i g(|\Psi|)\right]\Psi=0.\label{eq:genparaboliceqn}
\end{eqnarray}
Here, $\Psi\equiv\Psi(x,y,z)$, $(x,y)$ are Cartesian spatial coordinates, $z$ is an evolution parameter such as time or propagation distance, $\nabla_{\perp}\equiv(\partial/\partial x, \partial/\partial y)$ is the gradient operator in the $xy$ plane, $\alpha$ is a real number, and both $f(|\Psi|)$ and $g(|\Psi|)$
are real functions of a real variable. These last two functions respectively represent any non-dissipative and dissipative non-linearities which contribute to the equation of evolution. Special cases of the above class of equations are used to model a wide variety of both classical and quantum systems, such as monoenergetic electron beams \cite{AllenOxley2001}, beamlike monochromatic scalar electromagnetic waves \cite{SalehTeich1991}, intense scalar electromagnetic fields in nonlinear media \cite{Akhmediev1997}, uncharged superfluids \cite{Pismen1999} and (2+1)-dimensional Bose-Einstein condensates \cite{PitaevskiiStringari2003}. An important subclass of equation (\ref{eq:genparaboliceqn}) are those for which $f(|\Psi|)$ has its minimum value when $|\Psi|\ne 0$; in this case the field $\Psi(x,y,z)$ displays spontaneous symmetry breaking \cite{Pismen1999}.
\par
When studying the applicability of a particular form of Eq.~(\ref{eq:genparaboliceqn}) to a given physical system, one often encounters the following logic. (i) A particular functional form of the equation is postulated or derived; (ii) this equation is then used to predict the result of a given experiment or series of experiments; (iii) goodness-of-fit between the prediction and the data, often within the context of a parameterized model of the
particular experiment being undertaken, is used as an indicator of whether the given equation is satisfactory - if the equation is unsatisfactory, one modifies step (i) and repeats the cycle until satisfactory results are obtained \cite{Sabatier1990}.
\par
Notwithstanding the many successes of such a methodology, one may enquire if the process can be systematized. Specifically, can one determine (or ``measure") the equation of motion governing a complex field $\Psi$, when the only measurable quantity is $|\Psi|$?  In a novel approach, given in this paper, we give an affirmative answer to this question.
\par
A key assumption in our approach is that the desired equation belongs to the class of equations (\ref{eq:genparaboliceqn}). We start by obtaining a ``hydrodynamic'' formulation of this equation via the Madelung transformation \cite{Madelung1926}, $\Psi(x,y,z) = \sqrt{I(x,y,z)}\exp[i\Phi(x,y,z)]$, where $I(x,y,z)\equiv|\Psi(x,y,z)|^2$ is the probability density (or intensity, as appropriate) of the field and $\Phi(x,y,z)$ is its phase. Substitute this into Eq.~(\ref{eq:genparaboliceqn}), and then separate real and imaginary parts, to obtain:
\begin{eqnarray}
    \frac{\alpha}{2}\frac{\partial I}{\partial z} +
    \nabla_{\perp}\cdot\left(I\nabla_{\perp}\Phi\right) + I g(I)&=& 0,
    \label{eq:gentie3}\\
    \alpha\frac{\partial \Phi}{\partial z}
    -f(I) +|\nabla_{\perp}\Phi|^2
    - H(I,\nabla_{\perp}I) &=& 0, \label{eq:gentie4}
\end{eqnarray}
where the ``diffraction term'' $H(I,\nabla_{\perp} I)\equiv I^{-\frac{1}{2}}\nabla_{\perp}^2 \sqrt{I}$ is a measurable function of probability density or intensity (we will henceforth speak of intensity, rather than probability density, for convenience). The above pair of coupled equations is equivalent to the partial differential equation (\ref{eq:genparaboliceqn}) from which it was derived. Note that, for matter-wave fields, the
diffraction term disappears in the classical limit; for radiation wave-fields, this same term disappears in the geometric-optics limit.  For either case, the classical limit of Eq.~(\ref{eq:gentie4}) yields the Hamilton-Jacobi equation governing the evolution of the field.
\par
The continuity equation (\ref{eq:gentie3}) is independent of the functional form of the non-dissipative nonlinearity $f(I)$. Introducing the scaled phase $\tilde{\Phi}$ via $\tilde{\Phi}\equiv 2 \Phi / \alpha$, and restricting ourselves for the moment to the $g(I)=0$ (non-dissipative) case, this equation becomes:
\begin{equation}
\frac{\partial I}{\partial z} + \nabla_{\perp}\cdot
\left(I\nabla_{\perp}\tilde{\Phi}\right) =  0. \label{eq:gentie5}
\end{equation}
We now turn to the inverse problem of determining the phase of the field from intensity measurements alone. To this end we note the observation of Teague, made in the more limited context of complex fields that obey the special case of Eq.~(\ref{eq:genparaboliceqn}) in which both $f$ and $g$ are equal to zero, that the associated continuity equation (\ref{eq:gentie5}) may be used as a basis for deterministic phase retrieval \cite{Teague1983}. Specifically, if one measures both $I$ and $\partial I /\partial z$ over some plane $z= z_0$, with $I$ being strictly positive over some simply-connected region $\Omega$ in that plane, and zero outside $\Omega$, then a simple non-linear generalization \cite{Paganin2001} of a result due to Gureyev and Nugent
\cite{GureyevNugent1996} shows that Eq.~(\ref{eq:gentie5}) can be uniquely solved for the scaled phase $\tilde{\Phi}(x,y,z=z_0)$, up to an unknown additive constant. We stress that this reconstruction requires knowledge of neither $\alpha$ nor $f(I)$, relying rather on the hypothesis (which may be tested using the
self-consistency argument given near the end of this paper) that $\Psi$ obeys an otherwise unknown equation belonging to the class of equations (\ref{eq:genparaboliceqn}).
\par 
In retrieving the scaled phase $\tilde\Phi$, one is faced with the determination of the appropriate boundary conditions for $\tilde\Phi$. Here we briefly discuss the role of boundary conditions in the retrieval of $\tilde\Phi$. To be specific we consider a phase problem in optics. As has already been stated, knowledge of both $I$ and $\partial I /\partial z$ over a plane $z=z_0$, with $I$ being strictly positive over some simply-connected region in that plane, and zero outside, implies that the continuity equation (\ref{eq:gentie5}) (termed transport of intensity equation \cite{Teague1983}) can be uniquely solved for the scaled phase $\tilde{\Phi}(x,y,z=z_0)$ up to an unknown additive constant \cite{Paganin2001,GureyevNugent1996}. When solving the transport of intensity equation for the phase, boundary conditions may be treated explicitly (see e.g., \cite{Gureyev1995}), or implicitly, as in reference \cite{GureyevNugent1996}. Regarding the explicit treatment of boundary conditions, let us give a non-linear generalization of an argument due to Roddier \cite{Roddier1990} (see also \cite{Gureyev1995}), which shows how Neumann boundary conditions to the non-linear dissipative transport of intensity equation (\ref{eq:gentie3}) may be determined from modulus data alone. In this example, $z$ corresponds to a propagation distance, allowing us to place a thin black screen in the plane $z=z_0$, into which is cut a simply-connected convex aperture with smooth boundary $\partial\Omega$. We assume the experiment to be so arranged that the intensity of the field, in the plane of the aperture, is strictly positive within the aperture, and zero outside. Since the aperture is both smooth and convex, points on the boundary of the aperture may be specified by the single-valued continuous function $r=R(\theta)$, where $r$ and $\theta$ are radial and angular plane polar coordinates with respect to a given origin which lies within $\partial\Omega$. The intensity distribution, in the plane of the aperture, can then be written as $I(r,\theta,z_0)H[R(\theta)-r]$, where $H$ is the Heaviside step function, and $I(r,\theta,z_0)$ is everywhere smooth and continuous. Substituting into Eq. (\ref{eq:gentie3}), and writing in explicit functional dependencies for the sake of clarity, one obtains:
\begin{eqnarray}
\frac{\alpha}{2}\left\{\frac{\partial I(r,\theta,z)}{\partial
z}\right\}_{z=z_0}
=\delta[r-R(\theta)]\frac{\partial\Phi[R(\theta),\theta,z_0]}{\partial
n}I[R(\theta),\theta,z_0]\qquad\qquad\nonumber\\
-H[R(\theta)-r]\left\{\nabla_{\perp}\cdot\left[I(r,\theta,z_0)\nabla_{\perp}\Phi(r,\theta,z_0)\right]  +I(r,\theta,z_0)g[I(r,\theta,z_0)]\right\}. \qquad\label{eq:Neumann1}
\end{eqnarray}
Here, $\delta$ is the Dirac delta and $\partial/\partial n$
denotes differentiation along the outward-pointing normal to the
aperture edge. Evaluate the above expression at the edge of the
aperture; assuming the phase to be single-valued and continuous
within and on the aperture, and the intensity to be differentiable
within the aperture, the last term on the right side
of the above expression are negligible compared to the first
term. This leaves:
\begin{eqnarray}
\frac{\alpha}{2}\left\{\frac{\partial
I[R(\theta),\theta,z]}{\partial z}\right\}_{z=z_0}
=\delta[r-R(\theta)]\frac{\partial\Phi[R(\theta),\theta,z_0]}{\partial
n}I[R(\theta),\theta,z_0].  \label{eq:Neumann2}
\end{eqnarray}
This equation may be radially integrated, to yield the Neumann
boundary conditions which are required for the unique solution of
Eq. (\ref{eq:gentie3}) for the phase, up to an arbitrary additive
constant. From a physical point of view, we note that the singular behaviour of the
intensity derivative at the edge of the sharp
aperture is due to local energy flow transverse to the aperture
edge, as the field propagates from plane to plane.  Note, also,
that the above method for determining Neumann boundary conditions
requires knowledge of neither the non-dissipative non-linearity
$f$ nor the dissipative non-linearity $g$.
\par
Evidently, knowledge of $\tilde{\Phi}$ allows us to infer
$|\nabla_{\perp}\tilde{\Phi}|^2$ in Eq.~(\ref{eq:gentie4}).
Further, $\tilde{\Phi}$ can (up to an unknown additive constant
which, in general, differs from plane to plane) be determined over
two closely-spaced planes $z=z_0$ and $z=3z_0$, so that one may
infer $\partial\tilde{\Phi}/\partial z$ (on the $z = 2z_0$ plane) up to an unknown constant,
denoted by $R(z)$. Thus the estimated $z$-derivative of
$\tilde{\Phi}$ is $\partial \tilde{\phi}/\partial
z\equiv\partial\tilde{\Phi}/\partial z - R(2z_0)$, so that
Eq.~(\ref{eq:gentie4}) becomes:
\begin{equation}
    H(I,\nabla_{\perp}I) + f(I) -\frac{\alpha^2}{2}\left(
    \frac{\partial\tilde{\phi}}{\partial z} +R +
\frac{1}{2}\left|\nabla_{\perp}\tilde{\Phi}\right|^2\right) = 0.
\label{eq:gentie8}
\end{equation}
If we assume that the intensity is non-vanishing in $\Omega$, but
vanishes on its boundary, continuity of the intensity implies that
at least two points $(x_1,y_1)$ and $(x_2,y_2)$ in $\Omega$ can
have the same intensity.  For such pairs, $I_1 = I_2$, where
$I(x_j,y_j)\equiv I_j,j=1,2$. Evaluate Eq.~(\ref{eq:gentie8}) at
each point, and subtract the resulting two equations, to give:
\begin{equation}
\gamma_{12} - \alpha^2\delta_{12} = 0,
\label{eq:gamma_alpha_delta}
\end{equation}
where the known quantities $\gamma_{12}$ and $\delta_{12}$ are:
\begin{eqnarray}
    \gamma_{12} &\equiv& 2\left[H(I_1,\nabla_{\perp}I_1) - H(I_2,\nabla_{\perp}
    I_2)\right], \label{eq:gamma}\\
    \delta_{12} &\equiv& \frac{\partial(\tilde{\phi}_1-\tilde{\phi}_2)}
    {\partial z}+ \frac{1}{2}\left(\left|\nabla_{\perp}\tilde{\Phi}_{1}
    \right|^2-\left|\nabla_{\perp}\tilde{\Phi}_{2}\right|^2\right).
    \label{eq:delta}\qquad
\end{eqnarray}
Note that the unknown constant $R$ has been eliminated.
Equation~(\ref{eq:gamma_alpha_delta}) gives ${\alpha} =
\sqrt{{\gamma_{12}}/{\delta_{12}}}$, where a positive root is
chosen because, as can be easily shown by applying the divergence
theorem to Eq.~(\ref{eq:gentie3}) and then invoking conservation
of integrated intensity, $ \alpha
> 0$. Having inferred $\alpha$, one can now infer
the phase $\Phi(x,y,z=2z_0)$ of the wave-field via
$\Phi(x,y,z=2z_0)=\frac{1}{2}\alpha \tilde{\Phi}(x,y,z=2z_0)$. Consequently, and bearing
in mind that $\Psi$ is completely specified by its measured
modulus and inferred phase, we are able to reconstruct $\Psi$,
even though $\alpha$ and $f(I)$ (in the $g=0$ case of the
equation of motion (\ref{eq:genparaboliceqn})) are unknown.
\par
Having solved the inverse problem of reconstructing the wave-field
$\Psi$ from measurements of its intensity, we turn to a second
inverse problem which forms the core subject of the present note,
namely a means for inferring non-linear parabolic equations of
motion that govern a complex scalar field, given modulus
data alone. For the case where $g(I) = 0$, the only unknowns are $f(I)$ and an auxiliary constant
$R$. Substitute the expression for $\alpha$ into
Eq.~(\ref{eq:gentie8}) and rearrange, to obtain
\begin{equation}
     f(I)  = \frac{\gamma_{12}}{2\delta_{12}}\left(\frac{\partial
     \tilde{\phi}}{\partial z} +R + \frac{1}{2}\left|\nabla_{\perp}
     \tilde{\Phi}\right|^2\right)- H(I,\nabla_{\perp}I).
     \label{eq:f}
\end{equation}
With the exception of $R$, each term on the right hand side is a
known function of $x$ and $y$. Therefore, the functional form of
$f$ can only be measured up to an unknown constant. In many
practical cases, one knows {\em a priori} the value of $f(0)$, in
which case $f$ can be fully determined. Note that the constant
component of $f$ may always be removed, via the transformation
$\tilde{\Psi}\rightarrow\Psi\exp(i\kappa z)$, for suitable $\kappa$.
\par
Thus far we have considered the $g(I)=0$ case of
Eq.~(\ref{eq:genparaboliceqn}), which amounts to ignoring
dissipation. We now show one means by which dissipation can be
included in the analysis. Suppose that one prepares a $z$-directed
plane-wave state, for which $\Phi$ is independent of $x$ and $y$,
over the plane $z=2z_0$. For such a state, which may have intensity
modulations over the plane $z=2z_0$,
$\nabla_{\perp}\cdot(I\nabla_{\perp}\Phi)=0$. Thus
Eq.~(\ref{eq:gentie3}) becomes $\frac{1}{2}\alpha \partial I /
\partial z + I g(I)  =0$, therefore $g(I)/\alpha=-\frac{1}{2}
\partial \ln I / \partial z$. If one repeats this measurement for
a series of values of $I$, one can determine $g(I)/\alpha$, as a
function of $I$. Having determined this function, multiply both
sides of Eq.~(\ref{eq:gentie3}) by $2/\alpha$, and make use of the
fact that $\tilde{\Phi}\equiv 2 \Phi / \alpha$, to arrive at
$\partial I / \partial z + \nabla_{\perp}\cdot (I
\nabla_{\perp}\tilde{\Phi}) + 2 I g(I)/\alpha=0$.  Since the
function $g(I)/\alpha$ is now known, and both $I$ and $\partial I /
\partial z$ can be measured for a non plane-wave, the only unknown
in this equation is $\tilde{\Phi}$. This can be determined using
the method cited earlier in the text, replacing $\partial I /
\partial z$ with $\partial I / \partial z +  2 I g(I)/\alpha$.
Having so determined $\tilde{\Phi}$, both $\alpha$ and $f(I)$ can
be determined, using the same method as previously outlined.
\par
We now give a numerical example, to provide a simple demonstration of
the application of our methodology using simulated data. As an
initial condition for numerical modelling using
Eq.~(\ref{eq:genparaboliceqn}), we chose a zero-phase modulated
Gaussian wave-packet centered on $(x,y)=(x_0,y_0)$:
\begin{equation}
\Psi(x,y,z=0) = \sqrt{A\left(1+M_x\right)\left(1+M_y\right)}
e^{-\frac{1}{4}\left(\frac{r-r_0}{W}\right)^2}.\label{eq:gaussianintensity}
\end{equation}
Here, $A$ and $W$ are real parameters respectively denoting the
peak squared modulus and width of the packet,
$r_0~\equiv~\sqrt{x_0^2+y_0^2}$,
\begin{eqnarray}
M_x &\equiv& \delta e^{-\frac{1}{2}\left(\frac{r-r_0}{W}\right)^2}\cos\left[2\pi n (x-x_0)\right], \label{eq:modulationx}\\
M_y&\equiv& \delta
e^{-\frac{1}{2}\left(\frac{r-r_0}{W}\right)^2}\sin\left[2\pi
n(y-y_0)\right], \label{eq:modulationy}
\end{eqnarray}
and $\delta,n$ are real parameters. For the simulations
presented here, we chose $A = 10$, $W = 8$, $\delta = 0.01$ and
$n=20$; the position of the peak is located at $r_0 = 0.5$.
\par
In modelling the forward evolution of $\Psi$ from $z=0$ to $z>0$,
we must specify all terms in Eq.~(\ref{eq:genparaboliceqn}). We
choose $\alpha = 1$ for all simulations, which can always be
arranged if one rescales $z$ appropriately. The grid size is
$2025\times 2025$ pixels, with a $z$-step $\Delta z = 10^{-7}$,
run for $5\times 10^{3}$ time-steps. The squared modulus
of $\Psi$ (intensity) is measured every $100$ time-steps, so
that at the end of the simulations, we obtain $50$ measurements of
the intensity profile. The intensity is measured over the central
$1025\times 1025$ sub-image of the larger simulation grid,
corresponding to a spatial domain $[x]\times [y]$ of $[0,1]\times
[0,1]$. The spatial step of the simulation is therefore $\Delta h
= 1/1024$. The evolution of $\Psi(x,y,z=0)$ to $z>0$ is obtained
by finite difference discretisation of $\Psi$, with $\partial
\Psi/\partial z$ being approximated using fourth-order Runge-Kutta
differentiation.
\par
The phase retrieval step, which involves solving
Eq.~(\ref{eq:gentie5}) for $\tilde{\Phi}(x,y,z')$ given both $I$
and $\partial I/\partial z$ at a given value of $z=z'$, is
amenable to the full multi-grid method \cite{Mudpack}. Once
$\tilde{\Phi}$ and the derivative of $\tilde{\Phi}$ are obtained
from three slices of the measured intensity profile, we calculate
$\alpha$. For an intensity profile with maximum $I_m$, we infer
$\alpha$ at $0.2 I_m$, $0.4 I_m$, $0.6 I_m$ and $0.8 I_m$. For
each chosen intensity, we go through every point $(i,j)$ and check
if the chosen intensity lies between the intensity at $(i,j)$ and
$(i+1,j)$ or $(i,j)$ and $(i,j+1)$. If so, we store the point that
has the closest intensity to the chosen intensity. We use the
previously-described method to calculate $\alpha$ from each pair
of points, and thereby construct a histogram of $\alpha$. The
value of $\alpha$ is determined from the peak of the histogram,
with the error $\sigma_{\alpha}$ given by its full width at half
maximum.
\par
Once $\alpha$ has been determined, we calculate $\Phi$ from the
definition $\tilde{\Phi} = 2\Phi/\alpha$. The result is then
compared with the actual phase of the wave field $\Psi$ taken from
the simulations. As already stated, we can only measure phase up
to an arbitrary additive constant. For comparison, we shift
the phase so that its average value is zero in a given plane,
which corresponds to a particular global phase choice. The
accuracy of the numerical method for phase recovery is dependent on the resolution of
the numerical grid. It is found that the relative error in the
phase, namely the normalised RMS error between the retrieved and
the input phase, is of the order of $5\%$ on average. It is
expected that, since the multigrid method utilises smoothing and
coarse grid correction, the major source of error arises from
regions with large phase gradients. Further, since the phase is
undefined when the intensity vanishes, phase retrieval is expected
to be less accurate away from the interior region where the
intensity approaches zero.
\begin{table}
\caption{\label{tab:table1}Measured values of $\alpha$ and
$f(I)$, for the first series of simulations, which have $g(I)=0$. The
actual value of $\alpha$ is unity, with $f(I)$ as described
in the text. $\sigma_{f}$ denotes the RMS error in inferring
$f(I)$, after the constant offset has been adjusted (see text).}
%\begin{ruledtabular}
\begin{tabular}{ccc}
$\alpha$ & $f(I)$ & $\sigma_{f}$\\
\hline
$1.01\pm 0.02$                    & $-199.40+100.26 I$ & $0.26\%$\\
$1.00\pm 0.01$                    & $-52.85-10.03 I^2$ & $0.28\%$\\
$1.00\pm 0.02$         & $-297.17+401.19 I -40.12 I^2$ & $0.30\%$\\
$1.00\pm 0.01$ & $-105.57+20.30 I -10.07 I^2 +1.00 I^3$ & $0.66\%$\\
$1.01\pm 0.04$          & $-118.43+100.01\sin(\pi I)$      & $0.89\%$\\
\end{tabular}
%\end{ruledtabular}
\end{table}
\par
\begin{figure}[t]
\psfrag{a}{$\alpha$}
\psfrag{fI}{$f(I)$}
\psfrag{I}{$I$}

\psfrag{0.6}{$0.6$}
\psfrag{0.8}{$0.8$}
\psfrag{1.2}{$1.2$}
\psfrag{1.4}{$1.4$}

\psfrag{500}{$500$}
\psfrag{1000}{$1000$}
\psfrag{1500}{$1500$}
\psfrag{2000}{$2000$}

\psfrag{0.99}{$0.99$}
\psfrag{1}{$1$}
\psfrag{1.01}{$1.01$}
\psfrag{1.02}{$1.02$}
\psfrag{1.03}{$1.03$}

\psfrag{10}{$10$}
\psfrag{20}{$20$}
\psfrag{30}{$30$}
\psfrag{40}{$40$}
\psfrag{50}{$50$}

\psfrag{-200}{\hspace{-1.5mm}$-200$}
\psfrag{-150}{\hspace{-1.5mm}$-150$}
\psfrag{-100}{\hspace{-1.5mm}$-100$}
\psfrag{-50}{\hspace{-1.5mm}$-50$}

\psfrag{2}{$2$}
\psfrag{4}{$4$}
\psfrag{6}{$6$}
\psfrag{8}{$8$}
\psfrag{10}{$10$}

\begin{center}
        \resizebox{6cm}{4cm}{\includegraphics{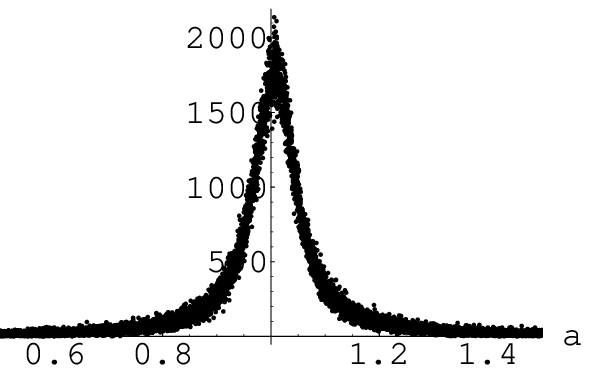}}
        \resizebox{6cm}{4cm}{\includegraphics{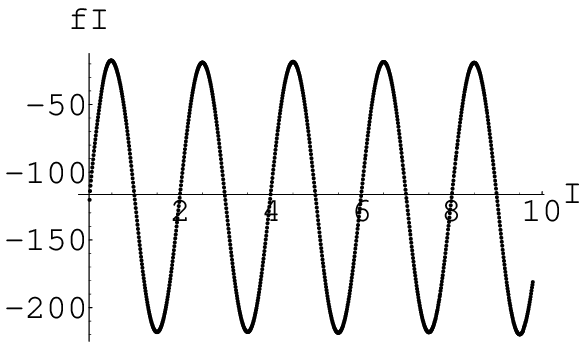}}\\
        \centering{(a)\hspace{6cm}(b)}
        \caption{\label{figure1}Measurements of (a) $\alpha$ and (b) $f(I)$ at the end of a simulation (based on the input $\alpha=1$, $f(I)=\sin(\pi I)$ and $g(I)=0$). The histogram of $\alpha$ shows a sharp peak, and this peak is considered to be the measured value (denoted as $\alpha_p$). Comparison with the input function $f(I)=\sin(\pi I)$ shows that we have accurately measured $f(I)$ up to a constant of $-118.43$. This constant part is due to the fact that we only measure the phase of $\Psi$ up to an arbitrary constant. Therefore this constant value may be used to correct our measurement of the phase of the wavefunction (see text).}
    \end{center}
\end{figure}

Obtaining $\alpha$ and $\tilde{\Phi}$ allows us to determine $f(I)$ using Eq.~(\ref{eq:f}). In the first series of simulations, which ignores dissipation, we used four different polynomials $f(I) = \gamma I + \kappa I^2 + \zeta I^3$ with ($\gamma,\kappa,\zeta) = (100,0,0)$, ($0,-10,0$), ($400,-40,0$) and ($20,-10,1$), and a sinusoidal function $f(I)=100\sin(\pi I)$. The recovered $\alpha$ and $f(I)$ are shown in Table~\ref{tab:table1}. This demonstrates accurate recovery of both $\alpha$ and $f(I)$ for nonlinear systems, including those with symmetry-breaking potentials. Measurement for the input $f(I)=100\sin(\pi I)$ ($\alpha = 1$) is shown in Fig.~\ref{figure1}. $\alpha$ can be read off from the peak of the histogram in (a). Accurate measurement of $\alpha$ leads to precise measurement of $f(I)$ up to a constant as shown in (b). This constant arises as result of $\Phi$ only being able to be retrieved up to an arbitrary constant, and this constant has been set to zero on both the first and the second planes. Since the constant part of $f(I)$ can be transformed into the phase of $\Psi$ as pointed out earlier, it may be used to appropriately adjust the constant part of the recovered phase on the second plane.
\begin{table}
\caption{\label{tab:table2}Input dissipation function, $g_i(I)$,
compared to recovered dissipation $g_r(I)$.  $\langle
g_r(I)\rangle$ is the recovered dissipation, averaged over 49
measurements, while $\sigma_{\langle g\rangle}$ is the RMS error in
this recovered dissipation.}
%\begin{ruledtabular}
\begin{tabular}{ccccc}
$g_i(I)$ & $g_r(I)$ & $\sigma_{g}$ & $\langle g_r(I)\rangle$ & $\sigma_{\langle g\rangle}$\\
\hline
$10 I$                    & $9.43 I^{1.04}$ & $2.11\%$      & $9.59 I^{1.03}$ & $1.30\%$\\
$2 I^2$                    & $2.22 I^{1.95}$ & $1.19\%$   & $2.00 I^{2.00}$ & $0.94\%$\\
\end{tabular}
%\end{ruledtabular}
\end{table}
\par
\begin{figure}[t]
\psfrag{gI}{$g(I)$}
\psfrag{I}{$I$}

\psfrag{2}{$2$}
\psfrag{4}{$4$}
\psfrag{6}{$6$}
\psfrag{8}{$8$}
\psfrag{10}{$10$}

\psfrag{-50}{\hspace{-1mm}$-50$}
\psfrag{50}{$50$}
\psfrag{100}{$100$}
\psfrag{150}{$150$}

\psfrag{g1}{$g_i(I)=10I$}
\psfrag{g2}{$g_i(I)=2I^2$}

\begin{center}
        \resizebox{6cm}{4cm}{\includegraphics{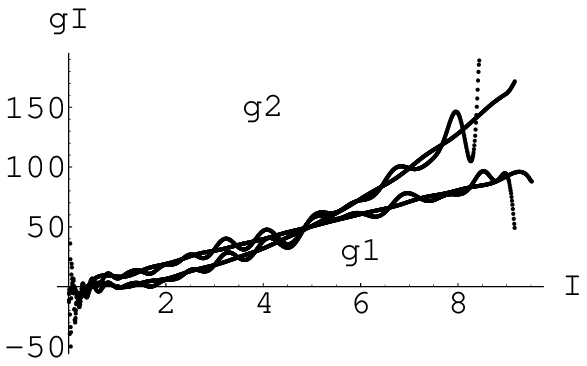}}
        \caption{\label{figure2}Measuring the nonlinear dissipation for the functions $g(I)=10I$ and $g(I)=2I^2$ ($\alpha=1$ and $f(I)=\sin(\pi I)$). Plots with more oscillations are obtained from a single measurement (from data at the end of the simulation), whereas the less oscillatory lines are taken from averaging over 49 measurements. This illustrates that the dissipation can be measured more accurately by averaging over larger samples. Averaging over many measurements may be a viable method for measuring nonlinear dissipations for highly fluctuating systems.}
    \end{center}
\end{figure}

The second series of simulations incorporate dissipation, which was assumed to be of the form $g(I) = D I^{\beta}$, where $D$ and $\beta$ are real numbers. The simulations were performed for various values of $D$ and $\beta$ with $\alpha = 1$ and $f(I) = 10\sin(\pi I)$. In this second series of simulations we show how dissipation can be measured, in addition to the other constants and functions which appear in Eq.~(\ref{eq:genparaboliceqn}). Previously we pointed out that the nonlinear dissipation term $g(I)$ can be measured using a plane wave formalism. For many systems it may be possible to construct a plane wave state for $\Psi$; for example, if $\Psi$ describes monoenergetic electron beams or an electromagnetic wave field. However, for the systems where $\Psi$ is a complex matter field describing an uncharged superfluid or a Bose-Einstein condensate, it may be difficult to construct a plane wave for $\Psi$. In this latter case the nonlinear dissipation $g(I)$ may be measured as follows. If the fluctuations of the system are small, the second term $\nabla_{\perp}\cdot(I\nabla_{\perp}\Phi)$ in Eq.~(\ref{eq:gentie3}) is expected to be much smaller than the first and last terms. In this case the nonlinear dissipation term can be approximated by 
\begin{equation}
	g(I) \approx -\frac{\alpha}{2I}\frac{\partial I}{\partial z}. \label{eq:measurediss}
\end{equation}
For a system with large fluctuations, we obtain the nonlinear dissipation by averaging the dissipation over many measurements. For $N$ measurements, Eq.~(\ref{eq:gentie3}) can be written as
\begin{equation}
\sum^N_{k=1}\left[\frac{\alpha}{2}\frac{\partial I}{\partial z} +
    \nabla_{\perp}\cdot\left(I\nabla_{\perp}\Phi\right) + I g(I)\right]_k= 0,
    \label{eq:gentie7}
\end{equation}
where $k$ denotes the $k$th measurement in the $z$-direction. The first and third terms in the square brackets in Eq.~(\ref{eq:gentie7}) are expected to increase with $N$, whereas the second (fluctuation) term is expected to approach zero in the limit of large $N$ (long term evolution), for a chaotic system. Therefore the nonlinear dissipation can be estimated using the average value 
\begin{equation}
	\langle g(I)\rangle \approx -\frac{\alpha}{2N}\sum^N_{k=1}\left[\frac{1}{I}\frac{\partial I}{\partial z}\right]_k. \label{eq:measurediss2}
\end{equation}
Note also that when measuring the nonlinear dissipation, the parameter $\alpha$ is not yet known. The nonlinear dissipation is in fact measured in terms of $\alpha$. This is not an issue since $g(I)/\alpha$ is the only quantity needed to measure other functions in the evolution equation. For convenience we use the notation $g(I)$ since we have set $\alpha = 1$. The results for two typical simulations are shown in Table~\ref{tab:table2}. The fourth column shows a more accurate determination of $g(I)$ when averaged over 49 measurements, which is an indication that the dissipations can be measured accurately for a system with large fluctuations (if a large number of measurements are obtained). Figure~\ref{figure2} shows the measured dissipation functions for $g_i(I)=2I^2$ and $g_i(I)=10I$. The dissipation function $g(I)$ is accurately recovered, with an accuracy that can be improved by averaging over many measurements.
\par
The methods for measuring the evolution equation presented in this letter can be tested on empirical data; in this context we point out its constraints. To precisely infer the equation of motion from modulus information, we require high-resolution data. With the chosen grid resolution, our method does not cope well with Poisson noise much larger than one part in ten thousand. This currently restricts its application to high resolution and
low-noise data. Notwithstanding this, there are currently many systems for which our method is applicable, such as the optics of intense electromagnetic beams in nonlinear media, and uncharged superfluid systems.
\par
We conclude by showing how the methods of this paper may be
extended to multi-component (2+1)-dimensional complex fields,
denoted by $\{\Psi_n(x,y,z)\}$, which comprise a set of $N$
complex scalar wavefunctions $\Psi_n\equiv \Psi_n(x,y,z), n=1,
\cdots, N$. This multi-component wavefunction might obey a system
of coupled nonlinear dissipative parabolic equations such as:
\begin{equation}
\left(i \alpha_n \frac{\partial}{\partial z} + \nabla_{\perp}^2
 + f_n + i g_n \right)\Psi_n=0,
\label{eq:nonlinearCoupledDE}
\end{equation}
where $\alpha_n$ are real numbers, while $f_n\left(I_1,\cdots,I_N\right)$ and $g_n\left(I_1,\cdots,I_N\right)$ are real functions of $N$ real variables, and $I_n\equiv |\Psi_n|^2$. The ``hydrodynamic'' formulation of Eq.~(\ref{eq:nonlinearCoupledDE}) is:
\begin{eqnarray}
 \frac{\partial I_n}{\partial z} +\nabla_{\perp} \cdot \left(I_n \nabla _{\perp} {\tilde{\Phi}}_n\right) + \frac{2 I g_n}{\alpha_n}&=& 0, \qquad\label{eq:multicomponentTIE}\\
  H_n +f_n -\frac{\alpha^2_n}{2} \left(\frac{\partial{\tilde{\Phi}}_n}{\partial z}+\frac{1}{2}|\nabla_{\perp}{\tilde{\Phi}}_n|^2\right)&=& 0, \label{eq:multicomponentEikonal}
\end{eqnarray}
where $H_n \equiv I_n^{-1/2} \nabla_{\perp}^2 \sqrt{I_n}$ is the generalized diffraction term.
Equation~(\ref{eq:multicomponentTIE}), which is uncoupled and
linear at the level of the unknown wavefunction phases $\Phi
_n\equiv \frac{1}{2}\alpha_n{\tilde{\Phi}}_n$, is independent of
$f_n$. When $g_n = 0$, this equation possesses a unique solution for the phases
${\tilde{\Phi}}_n$ (each up to an additive constant), given $I_n$
and $\partial I_n/\partial z$ as data, provided all such phases are
continuous. To subsequently solve Eq.~(\ref{eq:multicomponentEikonal}) for
$\alpha_n$ and $f_n$, our technique is to find pairs of points
with the same $f_n$. For arbitrary multi-component fields it is
not known how such pairs of points can be found (although such
pairs of points exist since $f_n$ vanishes on the boundary and is
non-vanishing in the interior). However, for two-component fields
($N=2$) in two spatial dimensions, we can always find such pairs
of points. That is we can construct a closed trajectory, $T_n$,
every point of which has the same intensity $I_n$ ($n=1$ say). As
we traverse a path in $T_1$, $I_2$ traverses the corresponding
path in $T_2$. However, since $T_1$ is a closed trajectory, $T_2$
is necessarily a closed trajectory. If these trajectories possess any intersection points, then $I_1=I_2$ at such points. It is then straightforward to follow the methods developed in this paper to infer the
equation of motion of the two-component field. A similar argument shows that in three spatial dimensions, it is
always possible to infer the equation of motion of a
three-component field.
\par
In order to measure equations of motion using the methods outlined
in this paper, we require that the modulus of each field component
be strictly positive over a simply-connected region $\Omega$.
This implies that these wavefunctions cannot possess topological
defects \cite{GureyevNugent1996,Pismen1999}.  However, once the
equation of motion has been inferred using a wavefunction free of
topological defects, wavefunction reconstruction in the presence
of defects may be performed using the method outlined by Tan {\em
et al.} \cite{Tan2003}.
\par
Finally, we emphasize that our method employs highly redundant systems of
equations, which yield multiple determinations for the desired
equation of motion.  If the postulated class of equations is
insufficiently large, this will be manifest as a lack of internal
consistency in the reconstructed equation, which thereby
constitutes a testable hypothesis rather than an assumption.

\ack We acknowledge support from the Australian Research Council
(ARC), the Victorian Partnership for Advanced Computing (VPAC),
and useful discussions with Y. Hancock.

\clearpage

\begin{thebibliography}{00}
\bibitem{Wheeler1982} J. A. Wheeler, Int. J. Theor. Phys. {\bf 21}, 555 (1982).
\bibitem{Nelles2001} O. Nelles, {\em Nonlinear System
Identification} (Springer, Berlin, 2001).
\bibitem{Voss1998} H. Voss, M. J. B\"{u}nner and M. Abel, Phys.
Rev. E {\bf 57}, 2820 (1998).
\bibitem{Bar1999} M. B\"{a}r, R. Hegger and H. Kantz, Phys.
Rev. E {\bf 59}, 337 (1999).
\bibitem{Gal2003} P. Le Gal, J. F. Ravoux, E. Floriani and T. Dudok
de Wit, Physica D {\bf 174}, 114 (2003).
\bibitem{Gabor1946}D. Gabor, J. Instn. Elect. Engrs, {\bf 93}, 429 (1946).
\bibitem{Pauli1933} W. Pauli, in {\em Handbuch der Physik}, edited by H. Geiger and K.
Scheel, Springer, Berlin (1933).
\bibitem{Aharonov1993} Y. Aharonov and L. Vaidman, Phys. Lett. A {\bf 178}, 38 (1993).
\bibitem{AllenOxley2001} L. J. Allen and M. P. Oxley, Opt. Commun. {\bf 199}, 65 (2001).
\bibitem{Paganin2001} D. M. Paganin and K. A. Nugent, {\em Non-Interferometric Phase Determination}, in P. W. Hawkes (ed.), {\em Advances in Imaging and Electron Physics}, Harcourt Publishers, Kent, {\bf 118}, 85 (2001).
\bibitem{Tan2003} Y-R. E. Tan, D. M. Paganin, R. P. Yu and M. J. Morgan, Phys. Rev. E {\bf 68}, 066602 (2003).
\bibitem{Born1993} M. Born and E. Wolf, {\em Principles of Optics} (Pergamon Press, Oxford, 1993), 6th ed.
\bibitem{SalehTeich1991} B. E. A. Saleh and M. C. Teich, {\em
Fundamentals of Photonics} (Wiley, New York, 1991).
\bibitem{Akhmediev1997} N. N. Akhmediev and A. Ankiewicz, {\em Solitons, Nonlinear Pulses and Beams} (Chapman and Hall, London, 1997).
\bibitem{Pismen1999} L. M. Pismen, {\em Vortices in Nonlinear Fields} (Oxford University Press, Oxford, 1999).
\bibitem{PitaevskiiStringari2003} L. Pitaevskii and S. Stringari,
{\em Bose-Einstein Condensation} (Oxford University Press, Oxford 2003).
\bibitem{Sabatier1990} P. C. Sabatier, {\em Modelling or Solving Inverse
Problems?}, in P. C. Sabatier (ed.), {\em Inverse Methods in Action}, Springer-Verlag,
Berlin, 1 (1990).
\bibitem{Madelung1926} E. Madelung, Z. Phys. {\bf 40}, 322 (1926).
\bibitem{Teague1983} M. R. Teague, J. Opt. Soc. Am. {\bf 73}, 1434
(1983).
\bibitem{GureyevNugent1996} T. E. Gureyev and K. A. Nugent, J. Opt. Soc. Am. A {\bf 13}, 1670 (1996).
\bibitem{Gureyev1995}T. E. Gureyev, A. Roberts and K. A. Nugent, J. Opt. Soc. Am. A {\bf 12}
1932 (1995).
\bibitem{Roddier1990} F. Roddier, Appl. Opt. {\bf 29} 1402 (1990).
\bibitem{Mudpack} http://www.scd.ucar.edu/css/software/mudpack

\end{thebibliography}
\end{document}